\documentclass[12pt]{article}
\title{The Froth of the Universe}
\author{Ll. Bel \\
Lab. Gravitation et Cosmology Relativistes. ESA 70 \\
{\small Tour 22-12, 4 place Jussieu, 75252 Paris}
}
\date{\today}
\begin{document}
\maketitle

\begin{abstract}
We consider a model of the Universe based on the equation of state
$p=(1/3)\rho (c/F)^2$, where $F$ is the scale factor. This model
behaves as an inflationary Universe from the beginning and during 
its early stages, and
behaves as dust matter during the stages of maximum expansion.
\end{abstract}

\section{Introduction}
Let us consider the Robertson-Walker line-element written using one of
its appropriate coordinates:

\begin{equation}
\label {1.1}
ds^2=-c^2dt^2+F^2(t)N^2(r)\delta_{ij}dx^idx^j, \quad
N(r)=1/(1+kr^2/4), \quad r=\sqrt{\delta_{kl}x^kx^l}
\end{equation}
to which we shall refer as the standard coordinates of the
Robertson-Walker metrics.
We shall assume that the scale factor $F$ is dimensionless, that
$t$ has dimensions of time $T$ and that $r$ has dimensions of space $L$. 
The curvature 
constant $k$ will have therefore dimensions $L^{-2}$ and the speed of
light $c$ dimensions $LT^{-1}$.

From both the geometrical and physical points of view the
line-element \ref{1.1} is strictly equivalent to the following one:

\begin{equation}
\label {1.2}
d\tau^2=dt^2-\frac{F^2(t)}{c^2}N^2(r)\delta_{ij}dx^idx^j,
\end{equation}
which can be written also in the following suggestive form:

\begin{equation}
\label {1.3}
d\tau^2=dt^2-\frac{1}{c_{eff}^2}N^2(r)\delta_{ij}dx^idx^j,
\end{equation}
where:

\begin{equation}
\label {1.4}
c_{eff}=c/F(t)
\end{equation}

For any local physical process whose evolution is fast compared
with the time evolution of the scale factor, and is also local in
space or $k=0$, the line-element \ref{1.3} is telling us that this process will
evolve as if the Universe were Minkowski space-time
with an effective speed of light given by \ref{1.4}. If the time scale
of the evolution of the system that is considered is not short, like
for instance if the system is the Universe itself, the conclusion
above does not follow from the form of the line-element \ref{1.3}, but
it is an heuristic idea\,\footnote{This idea was already mentioned in
\cite{AB}. It has recently been substantiated in different contexts. See for
example \cite{BM} and references therein} that in our opinion deserves to be 
analyzed.

More precisely, let $\rho(t)$ and $p(t)$ be the density and the
pressure of the fluid filling homogeneously and isotropically the
Universe. We are going to assume that in the phase during which the
Universe is dominated by radiation the equation of state is not $p=1/3\rho
c^2$ as it is usually assumed, but:

\begin{equation}
\label {1.5}
p=\frac{1}{3}\rho c^2/F^2=1/3\rho c_{eff}^2
\end{equation}

Section 2 will set up the basic equations that can be derived from
Einstein's equations and the equation of state above for any interval
of time where it can be relevant. In section 3 we shall justify this
equation of state from the point of view of Relativistic kinetic
theory by proposing an appropriate re-definition of the
energy-momentum tensor of a gas of zero mass particles. In section 4
we shall set up the equivalence of the equation of state \ref{1.5}
with a particular inflationary scalar field in the very early phase
of the evolution of the Universe. In section 5 we shall consider the
very simple model that follows from assuming this equation of state
all the way during the life-span of the Universe.

\section{The field equations}

For a Robertson-Walker line element, Einstein's field equations with
a perfect fluid energy-momentum source term are\,\footnote{There are
many good books on cosmology. Two of them are \cite{Weinberg} and 
\cite{Islam}}

\begin{equation}
\label {3.1}
S_{\alpha\beta}-\Lambda g_{\alpha\beta}=\frac{8\pi G}{c^2}T_{\alpha\beta}, \quad 
T_{\alpha\beta}=\rho u_\alpha u_\beta + p\hat g_{\alpha\beta}, \quad
\hat g_{\alpha\beta}=g_{\alpha\beta}+ u_\alpha u_\beta
\end{equation}
where $u^\alpha$ is the unit time-like vector of the world-lines of
the cosmological fluid. Its components in standard coordinates are:

\begin{equation}
\label {3.2}
u^0=1, \quad u^i=0
\end{equation}
$\rho$ is the mass density with dimensions $ML^{-3}$ and $p$ is the
pressure with dimensions $ML^{-1}T^{-2}$.

Eqs. \ref{3.1} can be reduced to the following two equations, a dot meaning a
derivative with respect to $t$:

\begin{equation}
\label {3.3a}
{\dot F}^2=1/3(8\pi G\rho +c^2\Lambda)F^2-c^2k
\end{equation}
\begin{equation}
\label {3.3b}
2F\ddot F+{\dot F}^2+kc^2=-\frac{8\pi G}{c^2}pF^2+c^2\Lambda F^2
\end{equation}
These two equations imply the conservation equation:

\begin{equation}
\label {3.4}
\dot \rho+3(\rho+p/c^2)\dot F/F=0
\end{equation}
which is also equivalent to:

\begin{equation}
\label {3.5}
\frac{d}{dF}(\rho F^3)=-(3p/c^2)F^2
\end{equation} 

We shall use the standard notation for the Hubble and the deceleration 
functions:
\begin{equation}
\label {3.6}
H=\dot F/F, \quad q=-F\ddot F/{\dot F}^2 
\end{equation}
and for a while we shall use units of length, time and mass such that:

\begin{equation}
\label {3.7}
c=8\pi G=1, \quad H_0 =1
\end{equation}
where $H_0$ is the present value of $H$. 

As it was said in the introductory section we are going to discuss a
model of the universe for which the equation of state \ref{1.5}
holds. The equation \ref{3.5} can be written as:

\begin{equation}
\label {3.8}
\frac{d}{dF}\ln \rho=-(F^{-2}+3)F^{-1}
\end{equation}
which integrated yields the $F$ dependence of $\rho$:

\begin{equation}
\label {3.9}
\rho=AF^{-3}\exp[1/(2F^2)]
\end{equation}
where $A$ is a constant of integration. The $t$ dependence of $F$ can
be inferred from the result of integrating Eq. \ref{3.3a}:

\begin{equation}
\label {3.10}
t(F_2)-t(F_1)=\int^{F_2}_{F_1}\frac{dF}{[1/3(\rho+ \Lambda)F^2-k]^{1/2}}
\end{equation}.

Eq. \ref{3.3a} can be re-written:

\begin{equation}
\label {3.11}
k=F^2(H^2-\Lambda/3)(\Omega-1)
\end{equation}
where as usual we have defined $\Omega$ as the quotient of $\rho$ and
the critical density $\rho_c$:  

\begin{equation}
\label {3.12}
\Omega=\rho/\rho_c, \quad  \rho_c=3H^2-\Lambda
\end{equation}

Eq. \ref{3.3b} becomes:

\begin{equation}
\label {3.14}
-2F^2qH^2+F^2H^2\Omega-1/3F^2\Lambda\Omega-2/3\Lambda F^2+H^2\Omega
-1/3\Lambda\Omega=0
\end{equation}
Solving the system of Eqs. \ref{3.12} and \ref{3.14} for $\Omega$ and
$\Lambda$ we get:

\begin{equation}
\label {3.15}
\Omega=\frac{2F^2\rho}{6F^2H^2(q+1)-\rho(F^2+1)}
\end{equation}
and:
\begin{equation}
\label {3.13}
\Lambda=-\frac{6F^2qH^2-\rho(F^2+1)}{2F^2}
\end{equation}

The two equations \ref{3.11} and \ref{3.13} would yield directly the
values of the two free parameters of the model derived from the
the density and the Hubble and deceleration functions at some
given time.

\section{Relativistic kinetic theory}

Let $f(x^\alpha, k_\beta)$ be the distribution function of a gas of
particles with mass $m=0$. The energy momentum tensor of such gas has always been defined
as\,\footnote{Relativistic kinetic
theory for a gas of massless particles filling the Universe at some
stage has been thoroughly studied in
\cite{EGS}. A gas of massive particles may behave also as a gas of
massless particles at some stages of the evolution of the Universe; 
see the preceding reference and 
\cite{Hakim}. See also \cite{Bel} for a discussion of the concept of
equilibrium in a cosmological context} 
: 

\begin{equation}
\label {2.1}
T_{\alpha\beta}(x^\mu) =\int_{C^+_x} f(x^\mu, k_\nu)k_\alpha k_\beta
\sqrt{-g}\frac{dk^1\wedge dk^2\wedge dk^3}{-k_0}
\end{equation}
where ${C^+_x}$ is the future-poynting light-cone with vertex at the
event $x$. 

A kinetic theory model of a domain of space-time filled with radiation
is then governed by a system of equations which are:

\begin{itemize}
\item Einstein's equations \ref{3.1} where $T_{\alpha\beta}$ is defined 
as above and
\item A kinetic equation for the distribution function $f(x^\alpha,
k_\beta)$. 
\end{itemize}
This equation is the Liouville equation if one assumes
either 1) that the radiation is so diluted that collisions can be
neglected or ii) that there is equilibrium balance. This
equation then reads:

\begin{equation}
\label {2.3}
Df\equiv k^\alpha\partial_\alpha f-
\Gamma^\alpha_{\lambda\mu}u^\lambda u^\mu
\frac{\partial f}{\partial u^\alpha}=0
\end{equation}
where $\Gamma^\alpha_{\lambda\mu}$ are the Christoffel symbols of the
line-element \ref{1.1}
In a cosmological Universe filled with black-body radiation these equations 
reduce to two groups of equations which using a system of standard
coordinates are:
\begin{itemize}
\item Einstein's equations \ref{3.3a}, \ref{3.3b} or/and \ref{3.4}  

where now $\rho$ and $p$ are given by:

\begin{equation} 
\label {2.4a}
\rho=4\pi\int_0^\infty f(t,\nu )\nu^3\, d\nu
\end{equation}
\begin{equation}
\label {2.4b}
p=4\pi\int_0^\infty f(t,\nu )\nu \,d\nu   
\end{equation}

where $f(t,\nu)$ is Planck's distribution function:

\begin{equation}
\label {2.5}
f=\frac{2h}{\exp((h\nu/k_BT(t))-1)}
\end{equation}
$k_B$ being Boltzman's constant and $T(t)$ the local temperature 
function:

\begin{equation}
\label {2.5bis}
T(t)=T_0F(t)
\end{equation}
$T_0$ being the temperature at some particular time.

\item The Liouville equation that in this particular situation reads:

\begin{equation}
\label {2.4c}
Df=\nu\partial_t f-H\nu^2\frac{\partial f}{\partial\nu}=0 
\end{equation}
\end{itemize}

Since $k^\alpha$ is a null vector the trace of the energy momentum
tensor \ref{2.1} is zero, $g^{\alpha\beta}T_{\alpha\beta}=0$, and this
is equivalent to the equation of state $p=1/3\rho$
and not to the equation of state \ref{1.5}. 

Let us remind that the world-lines of the cosmological fluid in a
Robertson-Walker universe are the world-lines of a conformal Killing
vector. That is to say that there exists a function $\xi(x^\alpha)$
such that the vector field $\xi^\alpha=\xi u^\alpha$ satisfies the
following equations:

\begin{equation}
\label {2.8}
\nabla_\alpha \xi_\beta+\nabla_\beta\xi_\alpha=2\Psi g_{\alpha\beta}
\quad \Psi=1/4\nabla_\sigma\xi^\sigma
\end{equation}

Consider the new definition of the energy-momentum tensor:

\begin{equation}
\label {2.7}
T_{\alpha\beta}(x^\sigma) =\int_{C^+_x} f(x^\mu, k_\nu)
\bar k_\alpha \bar k_\beta
\sqrt{-g}\frac{dk^1\wedge dk^2\wedge dk^3}{-k_0}
\end{equation}
where $\bar k_\mu$ is

\begin{equation}
\label {2.8bis}
\bar k_\mu=\xi^{-1}(k_\mu+(1-\xi^{-1})(k^\sigma u_\sigma) u_\mu)
\end{equation}
Consider now the following metric:

\begin{equation}
\label {2.9}
\bar g^{\alpha\beta}=g^{\alpha\beta}+(1-\xi^{-2})u^\alpha u^\beta
\end{equation}
We now have:

\begin{equation}
\label {2.10}
\bar g^{\alpha\beta}T_{\alpha\beta}=0
\end{equation}
and taking into account the fact that in standard coordinates we have
$\xi=F$ from Eq. \ref{2.10} it follows that the new expressions for $\rho$:

\begin{equation}
\label {2.11}
\rho(t)=aT^4(t), \quad a=\frac{8\pi^5k_B^4}{15h^3}
\end{equation}
and $p$ satisfy the equation of state \ref{1.5}. 

The fact that a small modification of the standard formalism of
Relativistic kinetic theory for a gas of zero mass particles allows 
to derive Eq. \ref{1.5} is, we believe, a sufficient justification for
this equation of state. Nevertheless one should be aware of an
important conceptual innovation. Contrary to what happened with
\ref{2.1}, the modified energy-momentum tensor \ref{2.7} will not
satisfy automatically the conservation equation \ref{3.4} as a
consequence of the Liouville equation \ref{2.3}. It will satisfy this
conservation equation only as a consequence of Einstein's equations
\ref{3.3a} and \ref{3.3b} and this means: i) that the Liouville equation
can not be implemented and ii) that the local temperature function
$T(t)$ in Planck's distribution function \ref{2.5} has to be derived by equating the
expression of the density above and that that we derived before in section 2, i.e.
Eq. \ref{3.9}:

\begin{equation}
\label {2.12}
AF^{-3}\exp[1/(2F^2)]=aT^4
\end{equation} 
This temperature function corresponds to a distribution function that 
satisfies a kinetic equation that can be written in many ways; a
possibility being:

\begin{equation}
\label {2.13}
Df(t,\nu)=1/8H(F^2-1)F^{-2}\nu h^{-1} f(f+2h)\ln(1+2hf^{-1})
\end{equation}

Therefore while we can still speak about a black-body radiation, the
equilibrium described by the distribution function \ref{2.5} is now
local and not global. This is a satisfactory situation for an expanding
Universe.
 
\section{Inflationary equivalent model}
It is well known
that it is not necessary to impose
that the source of a Robertson-Walker space-time is a perfect fluid
because this follows directly from Einstein's equations and the
assumptions that lead to the line-element \ref{1.1}. This is only
necessary if one wants to specify local conditions, like for example
choosing a particular equation of state. But by the same
token we can always assume that the source of a Robertson-Walker
universe is a scalar field  for which the energy-momentum tensor is:

\begin{equation}
\label {4.1}
T_{\alpha\beta}=\partial_\alpha\Phi\partial_\beta\Phi
-g_{\alpha\beta}[\partial_\mu\partial^\mu\Phi-V(\Phi)]
\end{equation}   
where $V(\Phi)$ is a potential function depending on the model being
considered. In standard coordinates 
the equivalence between a perfect fluid source and
a scalar field $\Phi(t)$ depending on the time only is set up by 
writing:

\begin{equation}
\label {4.2}
\rho=\frac12{\dot\Phi}^2+V(\Phi), \quad p=\frac12{\dot\Phi}^2-V(\Phi)
\end{equation}
equations that could be considered as the parametric representation
of an equation of state.

Let us assume that the model presented in the preceding sections is a
good description of the Universe during an early period of time for
which the scale factor $F$ is very small. It is an elementary
exercise to obtain the potential function $V(\Phi)$ that would make this model
look as a particular inflationary model. From \ref{4.2} and the above
equations we get:

\begin{equation}
\label {4.3}
\ddot\Phi+3H\dot\Phi+ V^\prime=0, \quad 
V^\prime=\frac{\partial V}{\partial\Phi} 
\end{equation}
Deriving the first Eq. \ref{4.2} with respect to time and using the
equation above we obtain:

\begin{equation}
\label {4.4}
{\dot\Phi}^2=-\frac{\dot\rho}{3H}
\end{equation}
which substituted in the first Eq. \ref{4.2} yields:

\begin{equation}
\label {4.5}
V=\rho+\frac{\dot\rho}{6H}
\end{equation}

Now we consider our particular solution characterized by a density
function \ref{3.9}. Deriving this function with respect to $t$ we obtain:

\begin{equation}
\label {4.6}
\dot\rho=-\frac{\rho}{F^3}(3F^2+1)\dot F
\end{equation} 
and substituting into \ref{4.4} and \ref{4.5} we get:

\begin{equation}
\label {4.7b}
{\dot\Phi}^2=\frac{\rho}{3F^2}(3F^2+1)
\end{equation}

\begin{equation}
\label {4.7a}
V=\frac{\rho}{6F^2}(3F^2-1)
\end{equation}
Let us consider the scalar field as a function $\Phi(F)$ of the scale
factor. Deriving this function with respect to $t$ and using Eqs.
\ref{4.7b} and \ref{3.3a} we obtain:

\begin{equation}
\label {4.8}
\frac{d\Phi}{dF}=\frac1F\sqrt{\frac{3F^2+1}{F^2+(\Lambda F^2-3k)/\rho)}}
\end{equation}
If in the interval under consideration $F$ is small enough 
then the r-h-t can be approximated to $1/F^2$ and
therefore we obtain: 

\begin{equation}
\label {4.10}
\Phi(F_2)-\Phi(F_1)=
\mp(\frac{1}{F_2}-\frac{1}{F_1})
\end{equation} 
and therefore, up to a sign factor and an arbitrary additive
constant, we get:

\begin{equation}
\label {4.11}
\Phi=\frac{1}{F}
\end{equation}  
in the largest interval for which the assumptions made are valid.

Using the expression above into \ref{4.7a} and keeping only the largest
term we obtain the inflationary
potential\,\footnote{Concerning the general ideas about inflation see
for instance \cite{Guth} or \cite{Islam}} that would be equivalent to the equation of state
considered in \ref{1.5}:

\begin{equation}
\label {4.12}
V(\Phi)=-\frac{A}{6}\Phi^5\exp[(1/2)\Phi^2]
\end{equation}

\section{A pure radiation very simple model}

Either from \ref{1.5} but still better from \ref{3.9} one can see that for 
large
values of $F(t)$ a model based on the equation of state
\ref{1.5} will be equivalent to a dust model, i.e. a matter model with
zero pressure. In fact under this condition the density 
behaves as:

\begin{equation}
\label {5.1}
\rho(F)=AF^{-3}, \quad
\end{equation}
It is therefore tempting to examine the properties of a Universe that
would satisfy the equation of state \ref{1.5} all the way during its
life-span. Such a model contains only two parameters: the
curvature constant $k$ and the cosmological constant $\Lambda$. We
shall determine them using the available theory presented before and
the present observable values of the Hubble constant $H_0$, the
deceleration parameter $q_0$ and
the present density. 
 
The point of view that we presented in the introductory section
raises the following important question: What is the constant $c$
appearing in the Robertson-Walker metric but also in Einstein's
constant, $8\pi G/c^2$, and in the energy-momentum tensor. It is
called the speed of light in vacuum\,\footnote{It should be called
otherwise because it is actually a property of space-time in the
sense that it belongs to all causal interactions} but we are also
saying here that since measuring the speed of light is a local
process the result of our measure is actually $c_{eff}$, not c. It
turns out then that for fixed units of length and time c can take any
value as long as we choose the present value of $F$ to be: 
$F_0=c/c_{eff}$. In particular we mankind, leaving for a very short
period of the history of the Universe, can choose to avoid the use of
two different quantities $c$ and $c_{eff}$ and decide to answer our
question by saying that $c$ is the measured value of the speed of
light in vacuum with the proviso that we take the present value of
$F$ to be:

\begin{equation}
\label {5.3}
F_0=1
\end{equation}
But remember this is not a choice of units condition as that that we made
in \ref{3.7}. This is a
conceptual decision.

From \ref{3.11} we have:

\begin{equation}
\label {5.5}
k=(1-\Lambda/3)(\Omega_0-1)
\end{equation}
and from \ref{3.15} and \ref{3.13} we get:

\begin{equation}
\label {5.6}
\Omega_0=\frac{2\rho_0}{6(q_0+1)-2\rho_0}
\end{equation}

\begin{equation}
\label {5.7}
\Lambda=-3q_0-\rho_0
\end{equation}
where $q_0$ and $\rho_0$ are the present observed values. 

We have now to make our mind about the present
density of the Universe.  To keep the model as simple as possible we
offer two possibilities:
	
i) To imagine that the gas of zero mass particles that we have
considered is in fact a faithful physical equivalent of the present
mixture of radiation and massive particles as if, so to speak, the latter 
were still diluted in the form of radiation. From this point of view
it would be safe to consider the present density, as a free parameter.

ii)To assume, as it has been repeatedly proposed, and as observations have
always consistently suggested\,\footnote{See for example \cite{Mandelbrot} and
references therein}, that the mass distribution in the
Universe is highly hierarchical. From this point of view there is no
objection in considering that the density of the mass distribution
could become much smaller than the density of the background
black-body radiation at some scale of distances as if, so to speak,
everything, from elementary particles to clusters of clusters of
clusters of galaxies, were nothing more than the {\it Froth of the
Universe}.

In both cases the present density and the present temperature
would be given by the Stefan-Boltzman law:
\begin{equation}
\label {5.8}
\rho_0=aT_0^4
\end{equation}
but the nicer aspect of the second choice is that for this case $T_0$
is a known quantity: the present temperatute of the background
black-body radiation. We give below a few numerical results
corresponding to this case.  
  
Using the following values: $q_0=.1$, $T_0$=2.7 K, and assuming
that $H_0^{-1}$ is $13$ billion years ($H_0=75\, \mbox{km/s/Mpc}$) we obtain 
using normalized units, i.e. those satisfying
the conditions \ref{3.7}:

\begin{equation}
\label {5.10}
\rho_0=1.3\times 10^{-4}, \quad \Omega_0=3.8\times 10^{-5}
\end{equation}
\begin{equation}
\label {5.9}
\Lambda=-.3, \quad k=-1.1 
\end{equation} 
Reverting to the MKS system of units these values correspond to the
following ones:

\begin{equation}
\label {5.11}
\rho_0=4.5\times 10^{-31}\, \mbox{kg/m}^3
\end{equation}

\begin{equation}
\label {5.12}
\Lambda=-2.\times 10^{-51} \mbox{m}^{-2}, \quad 
k=-7.3\times 10^{-51} \mbox{m}^{-2}
\end{equation}
Eq. \ref{3.10} with $F_1=0$ and $F_2=1$ gives the age of the Universe for this model:

\begin{equation}
\label {5.14}
\mbox{Age}=.77(H_0^{-1}), \hbox{      or       } 
\mbox{Age}=9.83\times 10^9\, \mbox{yr}  
\end{equation}

The maximum value $F_{max}$ of $F$ is given by the
root of the Eq. ${\dot F}^2=0$. Using the data above we have
$F_{max}=3.32$ and therefore Eq. \ref{3.10} with $F_1=0$ and $F_2=F_{max}$ gives half the life-span of the Universe, or:

\begin{equation}
\label{5.16}
\mbox{Life-span}=9.4, \hbox{     or     } 
\mbox{Life-span}=12.2\times 10^{10}\, \mbox{yr} 
\end{equation}
Notice that although $F_{max}$ is not a large number it is still
justified to claim that the energy density behaves as in \ref{5.1}
in the interval, say, $F=F_{max}\pm 1$.
\section{Concluding remarks}

This paper is based on three ideas. Namely:

\begin{itemize}
\item that the effective speed of light for any local process, in the
cosmological sense, is given by \ref{1.4}
\item that since the equation of state \ref{1.5} leads to an
inflationary behavior during the early stages of the Universe it
might deserve consideration, for the same reasons that one has for
other inflationary models, 

\item and that since this same equation of state is formally 
equivalent during
the latest stages of the evolution, to
assuming that the Universe is filled with dust matter, it is tempting
to consider a model for which this equation of state would be valid
from the beginning to the end.
\end{itemize}

We think that we have shown that these three ideas are connected. But
nothing forbids to consider them separately. We could for instance
keep the first one and forget the second and third. From this point
of view the single interest of the idea would be to predict a
variation of the effective speed of light given by:

\begin{equation}
\label {6.1}
\dot c_{eff}=-c_{eff}H_0
\end{equation}  
but it is doubtful that a local uncorrupted test of this prediction, of
cosmological origin, could be performed.

We could also keep the second idea and consider the Eq. \ref{1.5}
valid only during an early fraction of the history of the Universe to
be connected with other models at some other stages. And so on. Any
cosmological model can be improved making it more complicated. The
model of section 5 is just a model incorporating the three ideas
above in the simplest possible way. It has to be considered as a
template ready to be modified to fit other particular points of view.

\section*{Acknowledgements}
Many of the misprints and Latex errors of the initial manuscript were pointed out to me by A. Molina. Thanks Alfred !

\end{document}